\documentclass[twocolumn,epj]{epl2}

\usepackage{graphicx}
\usepackage{amsmath,amssymb}
\usepackage{bm}
\usepackage{color}
\usepackage{tikz}
\usepackage[colorlinks=true,linkcolor=blue,citecolor=blue,urlcolor=blue]{hyperref}

\usepackage[T1]{fontenc}
\usepackage[utf8]{inputenc}
\usepackage{booktabs}

\begin{document}

\title{Discrete Laplacian Structure and Kernel Reduction of the Gap Equation in $d=4k+3$ Gross--Neveu Model at Imaginary Chemical Potential}

\author{Evangelos G. Filothodoros\inst{1}}

\institute{%
 \inst{1} Department of Physics, Aristotle University, Thessaloniki, Greece
}

\eplabstract{We observe a remarkable mathematical structure in the gap equations of the large-$N$ Gross--Neveu model at imaginary chemical potential in odd spacetime dimensions $d = 4k+3$. We show they can be written as the sum of two parts: one defined by higher-order discrete Laplacian patterns and a cut-off dependent part given by truncated asymptotic expansion of a hypergeometric function. We argue that this picture corresponds to a deeper relationship between thermal field theories in these odd $d$ and exactly-solvable one-dimensional quantum problems. We find that the thermal mass at specific imaginary chemical potential values is fixed where internal energy balances entropic states of thermal modes which is physically equivalent with OPE inversion formula techniques where thermal mass values arise from transcendental sets of equations.}

\maketitle

\section{Introduction}

The $U(N)$ Gross-Neveu model in odd dimensions exhibits remarkable analytic properties at finite temperature. The phase structure of the model in $d$ Euclidean dimensions at imaginary chemical potential $\mu=-i\alpha$ and finite temperature has been studied extensively in thermal field theory and statistical transmutation \cite{Roberge,Neto,petkou2000,Filothodoros:2016txa, Filothodoros2018} but also it has been used at its regularised lattice version in order to achieve dynamical mass generation \cite{Roose}.At large $N$, the gap equation for the thermal mass $\sigma$ (with $z= e^{-\beta\sigma}e^{-i\beta\alpha}$) exhibits interesting regularities in odd dimensions $d = 3,7,11,15,\dots$, i.e., $d = 4k+3$.

We begin with the calculation of the two distinct parts of the fermion condensate gap equation. One vacuum part forms a truncated asymptotic expansion of a hypergeometric function (generating a hierarchy of UV-divergent terms with decreasing powers of cut-off $\Lambda$) and one thermal part governed by higher-order discrete Laplacian contributions. The coefficients of the thermal part are interestingly the same as those of the $(2k)$-th order finite difference operators---three-point stencils for $d=7$, five-point for $d=11$, seven-point for $d=15$, and so on.

We further observe a type of dimensional reduction: the $d$-dimensional thermal part of the thermal field theory seems equivalent to a one-dimensional quantum mechanical problem with Hamiltonian of non-interacting fermions. This correspondence arises from the thermal part of a $1d$ quantum system of two twisted fermionic harmonic oscillators. The main purpose of our results is not to solve the gap equation but rather to reveal its hidden discrete and low--dimensional origin.

\section{Large-$N$ Gap Equation at Imaginary Chemical Potential}

To calculate the condensate mass gap equation of the $U(N)$ Gross-Neveu model in the presence of imaginary chemical potential $\mu=-i\alpha$ (like the temporal part $A_0$ of a gauge field coupled to fermions \cite {Kapusta:2006pm}), we use the Euclidean action in $d$ dimensions
\begin{align}
\label{GNaction}
&S_{GN} = -\int_0^\beta \!\!\!dx^0\int \!\!d^{d-1}\bar{x}[\bar{\psi }^{i}(\slash\!\!\!\partial  -i\gamma_0\alpha)\psi ^{i}+\nonumber \\
&\frac{g_d}{2N}\left (\bar{\psi }^{i}\psi ^{i}\right )^{2} +i\alpha NQ_d]\,,\,\,\,i=1,2,..N\,.
\end{align}
where $Q_d$ is the eigenvalue of the $N$-normalized fermion number density operator $\hat{Q}_d=\psi^{a\dagger}\psi^a/N$ in $d$ dimensions and $\psi$ are $N$-component fermions. Introducing an auxiliary scalar field $\sigma$ by Hubbard--Stratonovich transformation and integrating out fermions in the large-$N$ limit, we obtain the canonical partition function
\begin{align}
\label{GNPF1}
&Z_f(\beta,Q_d)=\int({\cal D}\alpha)({\cal D}\sigma)e^{-N S_{d,eff}}\,,\\
\label{GNPF2}
&S_{d,eff}=iQ_d\int_0^\beta \!\!\!dx^0\!\!\int\!\!d^{d-1}\bar{x}\,\alpha -\frac{1}{2g_d}\int_0^\beta \!\!\!dx^0\!\!\int d^{d-1}\bar{x}\,\sigma^2+ \nonumber \\
&\rm Tr\ln\left(\slash\!\!\!\partial-i\gamma_0\alpha+\sigma\right)\,.
\end{align}
At temperature $T = 1/\beta$ and imaginary chemical potential $\mu = -i\alpha$:
\begin{align}
&\text{Tr}\ln\left(\slash\!\!\!\partial-i\gamma_0\alpha+\sigma\right) = \int \frac{d^{d-1}p}{(2\pi)^{d-1}}\Big[ \frac{E_p}{2} - \ln(1 + e^{-\beta E_p - i\beta\alpha}) \nonumber \\
&- \ln(1 + e^{-\beta E_p + i\beta\alpha})\Big]
\end{align}
where $E_p = \sqrt{p^2 + \sigma^2}$, $p = |\vec{p}|$.
Thus:
\begin{align}
&S_{d,eff} = -\frac{\sigma^2}{2g_d} +\int \frac{d^{d-1}p}{(2\pi)^{d-1}}\Big[ \frac{E_p}{2} - \frac{1}{\beta}\ln(1 + e^{-\beta E_p - i\beta\alpha})\nonumber \\
& - \frac{1}{\beta}\ln(1 + e^{-\beta E_p + i\beta\alpha})\Big]+(\text{charge part})
\end{align}
The gap equation $\partial S_{d,eff}/\partial\sigma = 0$ becomes

\begin{align}
&\frac{\sigma}{g_d} - \int \frac{d^{d-1}p}{(2\pi)^{d-1}}\Big[ \frac{\sigma}{E_p} - \frac{\sigma}{E_p}\frac{1}{e^{\beta E_p + i\beta\alpha} + 1} - \nonumber \\
& \frac{\sigma}{E_p}\frac{1}{e^{\beta E_p - i\beta\alpha} + 1}\Big] = 0
\end{align}
Recall that
\begin{equation}
\frac{1}{g_d}\equiv \frac{1}{g_*}+\frac{M_d\Gamma(1-\frac{d}{2})}{(4\pi)^{\frac{d}{2}}}
\end{equation}
where $M_d$ a mass scale separating weak from strong interaction regime, 
so the full gap equation is 
\begin{align}
&\frac{\sigma}{g_*}+\frac{\sigma M_d\Gamma(1-\frac{d}{2})}{(4\pi)^{\frac{d}{2}}} = \int \frac{d^{d-1}p}{(2\pi)^{d-1}}\Big[ \frac{\sigma}{E_p} - \frac{\sigma}{E_p}\frac{1}{e^{\beta E_p + i\beta\alpha} + 1} \nonumber \\
& -\frac{\sigma}{E_p}\frac{1}{e^{\beta E_p - i\beta\alpha} + 1}\Big]
\end{align}
Then
\begin{equation}
\frac{1}{g_d} =\frac{1}{g_*}+\frac{M_d\Gamma(1-\frac{d}{2})}{(4\pi)^{\frac{d}{2}}}= I_{\text{vac}} + I_{\text{therm}}
\end{equation}
where
\begin{equation}
I_{\text{vac}} = \int \frac{d^{d-1}p}{(2\pi)^{d-1}} \frac{1}{E_p}
\end{equation}
\begin{equation}
I_{\text{therm}} = -\int \frac{d^{d-1}p}{(2\pi)^{d-1}} \frac{1}{E_p} \left( \frac{1}{e^{\beta E_p + i\beta\alpha} + 1} + \frac{1}{e^{\beta E_p - i\beta\alpha} + 1} \right)
\end{equation}

\subsection{The thermal part}

The thermal part for $d = 4k+3$ is

\begin{align}
&I_{\text{therm}} = -C_d \int_0^\infty dp \, p^{4k+1} \frac{1}{\sqrt{p^2+\sigma^2}}\Big( \frac{1}{e^{\beta\sqrt{p^2+\sigma^2} + i\beta\alpha} + 1} + \nonumber \\
& \frac{1}{e^{\beta\sqrt{p^2+\sigma^2} - i\beta\alpha} + 1}\Big)
\end{align}
with $C_d = \dfrac{S_{4k+1}}{(2\pi)^{4k+2}}$.
Changing variables: $E = \sqrt{p^2+\sigma^2}$, so $p = \sqrt{E^2-\sigma^2}$ and $dp = \dfrac{E}{\sqrt{E^2-\sigma^2}} dE$.
Then:
\begin{align}
&p^{4k+1} dp = (E^2-\sigma^2)^{2k} \cdot \sqrt{E^2-\sigma^2} \cdot \frac{E}{\sqrt{E^2-\sigma^2}} dE = \nonumber \\
&(E^2-\sigma^2)^{2k} E dE
\end{align}
So:
\begin{equation}
\frac{1}{E_p} p^{4k+1} dp = (E^2-\sigma^2)^{2k} dE
\end{equation}
Thus:
\begin{equation}
I_{\text{therm}} = -C_d \int_\sigma^\infty dE \, (E^2-\sigma^2)^{2k} \left( \frac{1}{e^{\beta E + i\beta\alpha} + 1} + \frac{1}{e^{\beta E - i\beta\alpha} + 1} \right)
\end{equation}
If we expand $(E^2-\sigma^2)^{2k}$:
\begin{equation}
(E^2-\sigma^2)^{2k} = \sum_{j=0}^{2k} \binom{2k}{j} E^{4k-2j} (-\sigma^2)^j
\end{equation}
Thus:
\begin{align}
&I_{\text{therm}} =- C_d \sum_{j=0}^{2k} \binom{2k}{j} (-\sigma^2)^j \nonumber \\
&\int_\sigma^\infty dE \, E^{4k-2j}\Big( \frac{1}{e^{\beta E + i\beta\alpha} + 1} + \frac{1}{e^{\beta E - i\beta\alpha} + 1}\Big)
\end{align}
The binomial coefficients $\binom{2k}{j}$ are exactly the coefficients of the $(2k+1)$-point stencil for the $2k$-th derivative.

\begin{itemize}
\item
\textbf{For $d=7$ ($k=1$)}:
\begin{align}
&I_{\text{therm}}^{(7)} = -C_7\Big[ \int_\sigma^\infty dE \, E^4 f_{\text{therm}}(E) - \nonumber \\
& 2\sigma^2 \int_\sigma^\infty dE \, E^2 f_{\text{therm}}(E) + \sigma^4 \int_\sigma^\infty dE \, f_{\text{therm}}(E)\Big]
\end{align}
Coefficients: $(1, -2, 1)$ = 3-point stencil (2nd Laplacian).
\end{itemize}

\begin{itemize}
\item
\textbf{For $d=11$ ($k=2$)}:
\begin{align}
&I_{\text{therm}}^{(11)} = -C_{11}\Big[ \int_\sigma^\infty dE \, E^8 f_{\text{therm}}(E) - 4\sigma^2 \int_\sigma^\infty dE \, E^6 f_{\text{therm}}(E) + \nonumber \\
& 6\sigma^4 \int_\sigma^\infty dE \, E^4 f_{\text{therm}}(E) - 4\sigma^6 \int_\sigma^\infty dE \, E^2 f_{\text{therm}}(E)\nonumber \\
& + \sigma^8 \int_\sigma^\infty dE \, f_{\text{therm}}(E)\Big]
\end{align}
Coefficients: $(1, -4, 6, -4, 1)$ = 5-point stencil (4th Laplacian).
\end{itemize}

\begin{itemize}
\item
\textbf{For $d=15$ ($k=3$)}:
\begin{equation}
I_{\text{therm}}^{(15)} = -C_{15} \sum_{j=0}^{6} \binom{6}{j} (-\sigma^2)^j \int_\sigma^\infty dE \, E^{12-2j} f_{\text{therm}}(E)
\end{equation}
Coefficients: $(1, -6, 15, -20, 15, -6, 1)$ = 7-point stencil (6th Laplacian),
where $f_{\text{therm}}(E) = \dfrac{1}{e^{\beta E + i\beta\alpha} + 1} + \dfrac{1}{e^{\beta E - i\beta\alpha} + 1}$.
\end{itemize}

\subsection{The $\Lambda$ part}

The $\Lambda$ part for $d = 4k+3$ is
\begin{equation}
I_{\text{vac}} = \int \frac{d^{d-1}p}{(2\pi)^{d-1}} \frac{1}{E_p}
\end{equation}
Recall that 
\begin{equation}
\frac{1}{g_*}=\int^\Lambda \frac{d^dp}{(2\pi)^d} \frac{1}{p^2}
\end{equation}
For $d = 4k+3$, $d-2 = 4k+1$:
\begin{equation}
I_{\text{vac}} = \frac{S_{d-2}}{(2\pi)^{d-1}} \int_0^\Lambda dp \, p^{d-2} \frac{1}{\sqrt{p^2+\sigma^2}}
\end{equation}
with $S_{d-2} = \dfrac{2\pi^{(d-2)/2}}{\Gamma((d-2)/2)}$ and $\Lambda$ is the UV cutoff.
Since $p^{d-2} = p^{4k+1}$:
\begin{equation}
I_{\text{vac}} = C_d \int_0^\Lambda dp \, \frac{p^{4k+1}}{\sqrt{p^2+\sigma^2}}, \quad C_d = \frac{S_{4k+1}}{(2\pi)^{4k+2}}
\end{equation}
If we change variables: $p = \sigma t$, $dp = \sigma dt$:
\begin{equation}
I_{\text{vac}} = C_d \sigma^{4k+1} \int_0^{\Lambda/\sigma} dt \, \frac{t^{4k+1}}{\sqrt{1+t^2}}
\end{equation}
For large $\Lambda/\sigma$, we expand the integrand:
\begin{equation}
\frac{t^{4k+1}}{\sqrt{1+t^2}}= t^{4k} - \frac{1}{2}t^{4k-2} + \frac{3}{8}t^{4k-4} - \frac{5}{16}t^{4k-6} + \cdots
\end{equation}
So integrating all terms:
\begin{align}
&\int_0^{\Lambda/\sigma} t^{4k} dt = \frac{1}{4k+1} \left(\frac{\Lambda}{\sigma}\right)^{4k+1}\nonumber \\
&\int_0^{\Lambda/\sigma} t^{4k-2} dt = \frac{1}{4k-1} \left(\frac{\Lambda}{\sigma}\right)^{4k-1} \nonumber \\
&\int_0^{\Lambda/\sigma} t^{4k-4} dt = \frac{1}{4k-3} \left(\frac{\Lambda}{\sigma}\right)^{4k-3}
\end{align}
and so on.
Thus:
\begin{equation}
I_{\text{vac}} = C_d\Big[ \frac{\Lambda^{4k+1}}{(4k+1)} - \frac{1}{2}\frac{\Lambda^{4k-1}}{(4k-1)\sigma^{-2}} 
+ \frac{3}{8}\frac{\Lambda^{4k-3}}{(4k-3)\sigma^{-4}} - \cdots\Big]
\end{equation}
The coefficients $1, -\frac{1}{2}, \frac{3}{8}, -\frac{5}{16}, \dots$ are $\dfrac{(-1)^j(2j-1)!!}{2^j j!}$, which match the binomial expansion of $(1+x)^{-\frac{1}{2}}$.

\subsection{The full gap equation for $\sigma\neq 0$}

Since the first higher-powered term of $\Lambda^{4k+1}$ and the $\frac{1}{g_*}$ part cancel due to opposite signs, the full gap equation becomes

\begin{align}
&\frac{1}{g_*}+\frac{M_d\Gamma(1-\frac{d}{2})}{(4\pi)^{\frac{d}{2}}} = C_d \sigma^{4k+1} \int_0^{\Lambda/\sigma} dt \, \frac{t^{4k+1}}{\sqrt{1+t^2}}\nonumber \\ 
&- C_d \sum_{j=0}^{2k} \binom{2k}{j} (-\sigma^2)^j \int_\sigma^\infty dE \, E^{4k-2j} f_{\text{therm}}(E).
\end{align}

\section{Discrete Laplacian Structure - Dimensional ladder}

In $d=4k+3$ ($k=0,1,2,\dots$) the thermal part of the gap equation takes the form
\begin{align}
I_{\text{therm}}^{(d)}(\sigma)=C_d\int_{\sigma}^{\infty}
\bigl(E^{2}-\sigma^{2}\bigr)^{\frac{d-3}{2}}
f_{\text{therm}}(E)dE ,
\end{align}
with $\frac{d-3}{2}=2k$.  Defining the auxiliary dimension
\begin{align}
d'\equiv 2d-3 = 8k+3 ,
\end{align}
so that $\frac{d'-3}{2}=4k$, we observe that $I_{\text{therm}}^{(d)}$ can be obtained by applying a discrete higher‑derivative operator to $I_{\text{therm}}^{(d')}$.

Let $x=\sigma^{2}$ and consider the centred finite‑difference operator of order $2k$ with spacing $\epsilon$ \cite{Gattringer, Boyadzhiev}, which arises from the original forward difference operator, where the points are centered around x and it can be obtained by combining forward and backward differences,
\begin{align}
\Delta_{\epsilon}^{2k}F(x)=\sum_{j=0}^{2k}(-1)^{j}\binom{2k}{j}
F\!\bigl(x+(k-j)\epsilon\bigr).
\end{align}
Acting on $I_{\text{therm}}^{(d')}$,
\begin{align}
&\Delta_{\epsilon}^{2k}I_{\text{therm}}^{(d')}(x)=C_{d'}\sum_{j=0}^{2k}(-1)^{j}\binom{2k}{j}\nonumber \\
   &\int_{\sqrt{x+(k-j)\epsilon}}^{\infty}
   \bigl[E^{2}-(x+(k-j)\epsilon)\bigr]^{4k}f(E)\,dE .
\end{align}
Write each integral as $\int_{\sqrt{x}}^{\infty}$ minus (or plus) a boundary piece over the interval
$[\sqrt{x},\sqrt{x+(k-j)\epsilon}]$ (or $[\sqrt{x+(k-j)\epsilon},\sqrt{x}]$).  
For the part with common lower limit $\sqrt{x}$ we have, for fixed $E$,
\begin{align}
&\sum_{j=0}^{2k}(-1)^{j}\binom{2k}{j}
\bigl[E^{2}-(x+(k-j)\epsilon)\bigr]^{4k}\nonumber \\
 &= \epsilon^{2k}\frac{(4k)!}{(2k)!}\,\bigl(E^{2}-x\bigr)^{2k}+O(\epsilon^{2k+2}),
\end{align}
This happens because the left‑hand side is the $2k$‑th Laplacian operator of the degree‑$4k$ polynomial
$P_{4k}(u)=(E^{2}-u)^{4k}$, and
$\Delta_{\epsilon}^{2k}P_{4k}(x)=\epsilon^{2k}P_{4k}^{(2k)}(x)+O(\epsilon^{2k+2})$.
Since the boundary contributions are integrals over intervals with width $O(\epsilon)$, 
the integrand vanishes as $(E^{2}-x)^{4k}\sim O(\epsilon^{4k})$ at the edges,
so the total boundary contribution is $O(\epsilon^{4k+1})$, which should be considered negligible compared with
the $O(\epsilon^{2k})$ (since the bulk integral is finite due to the exponential decay of $f(E)$) for $k\ge1$. So:

\begin{equation}
\frac{\textbf{Boundary}}{\textbf{Bulk}}\rightarrow \epsilon^{2k+1}
\end{equation}
which goes to $0$ very fast for small $\epsilon$
and
\begin{align}
&\Delta_{\epsilon}^{2k}I_{\text{therm}}^{(d')}(x)
 =C_{d'}\epsilon^{2k}\frac{(4k)!}{(2k)!}\int_{\sqrt{x}}^{\infty}
   \bigl(E^{2}-x\bigr)^{2k}f(E)\,dE \nonumber \\
&+O(\epsilon^{2k+2}) .
\end{align}
But the right‑hand side is precisely
$\epsilon^{2k}\frac{C_{d'}}{C_d}\,I_{\text{therm}}^{(d)}(\sqrt{x})$.
Thus, up to higher orders in $\epsilon$,
\begin{align}
\boxed{\;
\Delta_{\epsilon}^{2k}\!\left[\frac{C_d}{C_{d'}\epsilon^{2k}}\frac{(2k)!}{(4k)!}
      I_{\text{therm}}^{(d')}(\sqrt{x})\right]
      = I_{\text{therm}}^{(d)}(\sqrt{x})
\;}.
\end{align}
The prefactor $C_d/C_{d'}\epsilon^{2k}$ carries a non‑trivial mass dimension. The extra mass powers could be replaced by
introducing a mass scale $\mu$ (e.g. the temperature $T$ since the free energy densities are analogue to $T^3$ at $d=3$, $T^7$ at $d=7$, etc.) and choosing the lattice spacing $\epsilon$ (in $\sigma^2$ space) to be $\epsilon=\mu^{2}$.  
With this choice,
\begin{align}
\frac{C_d}{C_{d'}\epsilon^{2k}}
   =\frac{C_d}{C_{d'}\mu^{4k}},
\end{align}
and the relation becomes
\begin{align}
\Delta_{\mu^{2}}^{2k}
\!\Bigl[\frac{C_d}{C_{d'}}\,\frac{(2k)!}{(4k)!}\mu^{-4k}
        I_{\text{therm}}^{(d')}(\sqrt{x})\Bigr]
 = I_{\text{therm}}^{(d)}(\sqrt{x}).
\end{align}
So we may write:
\begin{align}
\boxed{\;
I_{\text{therm}}^{(4k+3)}\;\propto\;
\Delta^{2k} I_{\text{therm}}^{(8k+3)}
\;},
\end{align}
where $\Delta^{2k}$ denotes the $2k$‑th order centred finite‑difference operator in the variable $\sigma^{2}$. 

We have to mention that although the discrete $\Delta^{2k}$ operator is familiar from lattice theory where these kind of operators act on spatial dimensions, here it acts on $x=\sigma^2$ space. This reveals a \emph{dimensional ladder}: the thermal
part of the gap equation in dimension $d=4k+3$ is obtained by applying a discrete
Laplacian of order $2k$ to the thermal part in the higher dimension $d'=2d-3=8k+3$.
The binomial coefficients that appear in the finite‑difference stencil are exactly
the same coefficients that multiply the integrals $I_{\text{therm}}$ in the explicit expansion
of $I_{\text{therm}}^{(d)}$, thereby explaining the origin of the discrete‑Laplacian
patterns observed in the Gross--Neveu gap equation. The construction fails for $d=4k+1$ since the polynomial kernel becomes odd and does not close under finite differentiation.

Explicitly \cite{LeVeque},
\begin{align}
\small
I_{7}(x)
&=
\frac{C_{7}}{C_{11}}
\frac{1}{12}
\lim_{\epsilon\to0}
\frac{
I_{11}(x+\epsilon)-2I_{11}(x)+I_{11}(x-\epsilon)
}{\epsilon^{2}},\\[6pt]
I_{11}(x)
&=
\frac{C_{11}}{C_{19}}
\frac{1}{720}
\lim_{\epsilon\to0}
\frac{1}{\epsilon^{4}}\big[
I_{19}(x+2\epsilon)-4I_{19}(x+\epsilon) \nonumber \\
&+6I_{19}(x)-4I_{19}(x-\epsilon)+I_{19}(x-2\epsilon)
\big]
\end{align}

\section{The fixed cut-off and $d-2$ free energy}

Following polylogarithm identities from \cite{Zagier1, Zagier2} we take useful results at $d=4k+3$, $k=1,2,3$

\begin{itemize}
\item{$d=7$}
\begin{equation}
D_5(z) = -\frac{\ln^2|z|}{6} D_3(z) - \frac{\ln^4|z|}{24} D_1(z) + R_5(z)
\end{equation}
\end{itemize}

\begin{itemize}
\item{$d=11$}
\begin{align}
&D_9(z) = -\frac{\ln^2|z|}{14} D_7(z) - \frac{\ln^4|z|}{280} D_5(z)-\nonumber \\
&\frac{\ln^6|z|}{5040} D_3(z) -\frac{\ln^8|z|}{40320} D_1(z)+ R_9(z)
\end{align}
\end{itemize}

\begin{itemize}
\item{$d=15$}
\begin{align}
&D_{13}(z) = -\frac{\ln^2|z|}{22} D_{11}(z) - \frac{\ln^4|z|}{792} D_{9}(z)-\frac{\ln^6|z|}{33264} D_7(z) \nonumber \\
&-\frac{\ln^8|z|}{1330560} D_5(z)-\frac{\ln^{10}|z|}{39916800} D_3(z) -\frac{\ln^{12}|z|}{95800320} D_1(z)+\nonumber \\
& R_{13}(z)
\end{align}
\end{itemize} 
From the gap equation of the model at $d=7$ \cite{Filothodoros2018}
\begin{align}
\label{gap71}
&\sigma\Big[-M_7\beta^5+D_5(-z)+\frac{1}{6}\ln^2\!|z|\left(D_3(-z)+\frac{\beta^3\Lambda_7^3}{45\pi}\right)+\nonumber \\
&+\frac{1}{24}\ln^4\!|z|\left(D_1(-z)-\frac{\beta\Lambda_7}{5\pi}\right)\Big]=0
\end{align}
we take that it reduces at the critical point $M_7=0$ to
\begin{equation}
R_5(z)+\text{$\beta\Lambda_7$\,(terms)}=0
\end{equation}
with
\begin{equation}
R_5(z) = -\operatorname{Li}_5(z) + \ln|z| \operatorname{Li}_4(z) - \frac{\ln^2|z|}{3} \operatorname{Li}_3(z)- \frac{ \ln^5|z|}{90}
\end{equation}
where $\mu=-i\pi$ is at the middle point of bosonization thermal window examined in \cite{Filothodoros:2016txa, Filothodoros2018} and $z=e^{-x}$. Then if $R_5$ function is zero, we obtain with a short excursion in Mathematica
\begin{equation}
R_5(e^{-x})=0 \rightarrow x=2.17755
\end{equation}
\begin{align}
&R_9(z) = -\operatorname{Li}_9(z) + \ln|z| \operatorname{Li}_8(z) - \frac{3\ln^2|z|}{7} \operatorname{Li}_7(z)+ \nonumber \\
&\frac{2\ln^3|z|}{21} \operatorname{Li}_6(z)-\frac{\ln^4|z|}{105} \operatorname{Li}_5(z)-\frac{\ln^9|z|}{198449}
\end{align}
When 
\begin{equation}
R_9(e^{-x})=0\rightarrow x=3.55044
\end{equation}
and
\begin{align}
&R_{13}(z) = -\operatorname{Li}_{13}(z) + \ln|z| \operatorname{Li}_{12}(z) - \frac{5\ln^2|z|}{11} \operatorname{Li}_{11}(z)+ \nonumber \\
&\frac{4\ln^3|z|}{33} \operatorname{Li}_{10}(z) -\frac{2\ln^4|z|}{99} \operatorname{Li}_9(z)+\frac{\ln^5|z|}{495} \operatorname{Li}_8(z)-\nonumber \\
&\frac{\ln^{13}|z|}{2872670000}
\end{align}
When 
\begin{equation}
R_{13}(e^{-x})=0\rightarrow x=4.93245
\end{equation}
and so the gap equation of the model reduces at the critical point $M_d=0$ to
\begin{equation}
R_{d-2}(z)+\text{$\beta\Lambda_d$\, (terms)}=0
\end{equation}
In the fermionic gap equation of $d$ dimensions we may say that it is mathematically reduced to a $(d-2)$-dimensional renormalized free energy which vanishes at a specific thermal mass $\beta\sigma=x$.
This condition dynamically fixes the UV cutoff $\Lambda$:
\begin{equation}
F_{d-2}(\sigma)=0.
\end{equation}
Thus, the $R(\sigma)$ functions that survive the binomial cancellations are the effective
thermal free energies of dimensionally reduced theories in $d-2$ dimensions:
\begin{equation}
\boxed{
F_d^{\mathrm{eff}}(\sigma)=F_{d-2}^{\mathrm{thermal}}(\sigma).
}
\end{equation}
Taking the derivative of $F_d$ with respect to $\sigma^2$ yields
\begin{equation}
\frac{\partial F_d(\sigma)}{\partial \sigma^2}
=
\frac{1}{2}
\int \frac{d^{d-1}p}{(2\pi)^{d-1}}
\frac{1}{E_p}
\Big[1-2f(E_p)\Big]
\;\propto\;
F_{d-2}^{\mathrm{thermal}}(\sigma)
\end{equation}
Thus the gap equation is the derivative of the free energy with respect to the mass parameter.

\begin{equation}
\boxed{
\text{Gap equation in } d
=
\frac{\partial F_d}{\partial \sigma^2}
\sim
F_{d-2}
}
\end{equation}
The operator $\frac{\partial}{\partial \sigma^2}$ acting on free energy has a crucial role in \cite{Karydas} when acting on the free energy of a quantum mechanical system to obtain one point functions.
At finite temperature the free energy equals
\begin{equation}
F = U - TS,
\end{equation}
where $U$ is the internal energy and $S$ the entropy.
Thus the above condition is equivalent to
\begin{equation}
\boxed{
U_{d-2}(\sigma)=T\,S_{d-2}(\sigma).
}
\end{equation}
Hence, the thermal mass $x=\beta\sigma$ is fixed and calculated at the point where the energy cost of creating massive
excitations is exactly balanced by the entropic gain from populating thermal modes. The vanishing of the $R(x)$ function after binomial cancellation is a physical observation where in $d=4k+3$, the large-$N$ Gross--Neveu model at finite temperature has a point where all UV vacuum contributions give a specific low boundary of cut-off energy. The $d-2$ reduction represents a thermodynamic balance of an effective lower-dimensional theory, where the thermal part of the gap equation scales like the free energy of a $d-2$ model. The equilibrium condition in the high-$d$ theory is exactly equivalent to a thermodynamic balance condition in a lower-$d-2$ theory. Thus although $F_{d-2}$ is kinematic in origin, it has real physical meaning.

\paragraph{For $d=7$, $k=1$:}

At the critical point $M_7=0$ of the theory where the mass scale $M_7$ separates strong from weak coupling regime, the gap equation becomes
 
\begin{equation}
-\frac{\Lambda_7^3}{6}+\frac{3\Lambda_7 |\ln(e^{-x})|}{8}=R_5(e^{-x})
\end{equation}

At $x_7=2.17755$, $R_5(x_7)=0$, so
the first trivial solution is $\Lambda_7=0$ at $x_7$ and the non-trivial solution is
\begin{equation}
-\frac{\Lambda_7^3}{6}+\frac{3\Lambda_7 |\ln(e^{-x_7})|}{8}=0\rightarrow \Lambda_7=\frac{3}{2}\sqrt{|\ln(e^{-x_7})|}
\end{equation}

\paragraph{For $d=11$, $k=2$:}
At the critical point $M_{11}=0$ of the theory where the mass scale $M_{11}$ separates strong from weak coupling regime, the gap equation becomes
\begin{align}
&\frac{\Lambda_{11}^7}{14}-\frac{3\Lambda_{11}^5\ln^2(e^{-x})}{40}+\frac{5\Lambda_{11}^3\ln^4(e^{-x})}{48}-\frac{35\Lambda_{11}\ln^6(e^{-x})}{128}=\nonumber \\
&R_9(e^{-x})
\end{align}
At $x_{11}=3.55044$, $R_9(x_{11})=0$, so
the first trivial solution is $\Lambda_{11}=0$ at $x_{11}$ and the non-trivial solution is
\begin{align}
&\frac{\Lambda_{11}^7}{14}-\frac{3\Lambda_{11}^5\ln^2(e^{-x_{11}})}{40}+\frac{5\Lambda_{11}^3\ln^4(e^{-x_{11}})}{48}-\nonumber \\
&\frac{35\Lambda_{11}\ln^6(e^{-x_{11}})}{128}=0\rightarrow \Lambda_{11}=\frac{5}{2}|\ln(e^{-x_{11}})|
\end{align}
The pattern continues for $d=15,19,\dots,4k+3$ with general form

\begin{equation}
\boxed{
\Lambda_{4k+3}=\frac{2k+1}{2}|\ln(e^{-x_{4k+3}})|^{\frac{k}{2}}}
\end{equation}
where $x_{4k+3}$ solves $R_{4k+1}(e^{-x})=0$.

The condition $F_{d-2}(\sigma)=0$ has an analogue picture with the OPE inversion techniques developed by \cite{Karydas,david2025,SingleVP, stergiou2018, kumar2023}. In their formalism non-trivial values of thermal mass arise from transcendental equation that ensures the absence of an infinite set of operators from the spectrum. In our case, equation $R_{d-2}(x)=0$ is doing the same work by balancing the UV-parts against finite thermal contributions arising from Bloch-Wigner-Ramakrishnan identities. As a result, the thermal contribution to the gap equation behaves as if governed by bosonic statistics, while the energy spectrum remains that of fermionic excitations. For smaller values of $\sigma$, entropy dominates due to the increasing number of low-energy modes, whereas for larger $\sigma$, the internal energy suppresses thermal occupation. We may argue that the solution of the gap equation selects the unique value of $\sigma$ at which these competing effects are equal. So, the variational method of large-$N$ effective action and the OPE inversion are two different sides of the same underlying physics that emerge a finite, thermal mass at criticality.

\section{Mapping to $1d$ quantum theory}

\subsection{The fermionic model in $d=3$}

For dimension $d=3$ \cite{petkou2000} (corresponding to $k=0$ or $d=4k+3$), 
the general formula for the thermal part simplifies to
\begin{equation}
I_{\text{therm}}^{(3)}(\sigma)=-C_3\int_{\sigma}^{\infty}f(E)\,dE ,
\end{equation}
The constant $C_3=1/(2\pi)$.
The integral can be evaluated by setting $x=\beta E$, $dx=\beta dE$:
\begin{align}
I_{\text{therm}}^{(3)}(\sigma)
 &=-\frac{1}{2\pi\beta}\int_{\beta\sigma}^{\infty}
   \Bigl(\frac{1}{e^{x+i\beta\alpha}+1}
         +\frac{1}{e^{x-i\beta\alpha}+1}\Bigr)\,dx .
\end{align}
Using the identity
$\displaystyle\int_{a}^{\infty}\frac{dx}{e^{x+c}+1}=\ln\!\bigl(1+e^{-a-c}\bigr)$, we obtain
\begin{align}
I_{\text{therm}}^{(3)}(\sigma)
 &=-\frac{1}{2\pi\beta}\Bigl[
      \ln\!\bigl(1+e^{-\beta\sigma-i\beta\alpha}\bigr)
     +\ln\!\bigl(1+e^{-\beta\sigma+i\beta\alpha}\bigr)\Bigr] \nonumber\\
 &=-\frac{1}{2\pi\beta}\,
   \ln\!\Bigl[1+2e^{-\beta\sigma}\cos(\beta\alpha)+e^{-2\beta\sigma}\Bigr].
\end{align}

Introducing the complex variable $z=e^{-\beta\sigma-i\beta\alpha}$ (so that
$\bar{z}=e^{-\beta\sigma+i\beta\alpha}$) the result takes the form
\begin{align}
\boxed{\;
I_{\text{therm}}^{(3)}(\sigma)
      =-\frac{1}{2\pi\beta}\bigl[\ln(1+z)+\ln(1+\bar{z})\bigr]
\;}.
\end{align}

\subsection{Free energy for 1D twisted harmonic oscillators with fermionic statistics}

We consider a quantum system of 1D twisted harmonic oscillators with fermionic statistics. Each pair of fermions has two species $\alpha = 1,2$ with operators $f_{n,\alpha}$, $f_{n,\alpha}^\dagger$ satisfying \cite{Vernier}
\begin{align}
\{ f_{n,\alpha}, f_{m,\beta}^\dagger \} = \delta_{nm} \delta_{\alpha\beta}, \quad
\{ f_{n,\alpha}, f_{m,\beta} \} = 0.
\end{align}
The $U(1)$ charge operator is defined as
\begin{align}
Q = \sum_{n=1}^N \big( N_{n,1} - N_{n,2} \big), \qquad 
N_{n,\alpha} = f_{n,\alpha}^\dagger f_{n,\alpha}.
\end{align}
We compute the twisted partition function
\begin{equation}
Z = \operatorname{Tr} e^{-\beta H + i\beta \alpha Q},
\end{equation}
where $i\alpha$ is the chemical potential conjugate to $Q$.

For non-interacting fermions ($k=0$) with mass $m$, the Hamiltonian is
\[
H_0 = \sum_{n=1}^N \sum_{\alpha=1}^2 m f_{n,\alpha}^\dagger f_{n,\alpha}.
\]
Since the system is translationally invariant, we Fourier transform:
\[
f_{n,\alpha} = \frac{1}{\sqrt{N}} \sum_{q} e^{iqn} \tilde{f}_{q,\alpha}, \quad
q = \frac{2\pi\ell}{N},\ \ell = 0,\dots,N-1.
\]
Then
\[
H_0 = \sum_{q} m \big( \tilde{f}_{q,1}^\dagger \tilde{f}_{q,1} + \tilde{f}_{q,2}^\dagger \tilde{f}_{q,2} \big).
\]
The charge in momentum space is
\[
Q = \sum_q \big( \tilde{f}_{q,1}^\dagger \tilde{f}_{q,1} - \tilde{f}_{q,2}^\dagger \tilde{f}_{q,2} \big).
\]
Thus the exponent becomes
\begin{equation}
-\beta H_0 + i\beta\alpha Q = \sum_q \big[ -\beta(m - i\alpha) \tilde{f}_{q,1}^\dagger \tilde{f}_{q,1} 
- \beta(m + i\alpha) \tilde{f}_{q,2}^\dagger \tilde{f}_{q,2} \big].
\end{equation}
For each fermionic mode, the trace over the two-dimensional Hilbert space ($n_q=0,1$) gives
\begin{equation}
\operatorname{Tr} e^{-\beta E N_q} = 1 + e^{-\beta E}.
\end{equation}
Therefore,
\begin{equation}
Z = \prod_q \big[ 1 + e^{-\beta(m - i\alpha)} \big] \big[ 1 + e^{-\beta(m + i\alpha)} \big].
\end{equation}
The free energy $F = -\beta^{-1} \ln Z$ is
\begin{equation}
F = -\frac{1}{\beta} \sum_q \Big[ \ln\!\big(1 + e^{-\beta(m - i\alpha)}\big) 
+ \ln\!\big(1 + e^{-\beta(m + i\alpha)}\big) \Big].
\end{equation}
In the thermodynamic limit $N \to \infty$, $\frac{1}{N}\sum_q \to \frac{1}{2\pi}\int_{-\pi}^{\pi} dq$, yielding
\begin{align}
\frac{F}{N} &= -\frac{1}{2\pi\beta} \int_{-\pi}^{\pi} 
\Big[ \ln\!\big(1 + e^{-\beta(m - i\alpha)}\big) 
+ \ln\!\big(1 + e^{-\beta(m + i\alpha)}\big) \Big] dq \nonumber \\
I_0 &= -\frac{1}{\beta} \Big[ \ln\!\big(1 + e^{-\beta(m - i\alpha)}\big) 
+ \ln\!\big(1 + e^{-\beta(m + i\alpha)}\big) \Big]= \nonumber \\
&2\pi I_{\text{therm}}^{(3)}(\sigma)
\end{align}

Now, it is easy to show that the thermal part $I_0$ of the $1d$ fermionic system generates the higher dimensional theory thermal part through repeated integration as a building block for higher dimensional thermal parts. In order to do this we define an operator $\mathcal{I}$ that acts on a function $\mathcal{F(\sigma)}$ as

\begin{equation}
\mathcal{I}^n[\mathcal{F(\sigma)}]
\end{equation}
So if one want to extract all thermal kernels he has to continue integration.
Using the integration identity \begin{equation}
\label{RepIint}
\mathcal{I}^{n}[F_0](\sigma)
=
\frac{1}{2^{n} n!}
\int_{\sigma}^{\infty}
(E^{2}-\sigma^{2})^{n} f(E)\, dE
\end{equation}
with $n=2k$ or since $\frac{d-3}{2}=2k$ for $d=4k+3$, we obtain
\begin{equation}
\mathcal{I}^{2k}[I_0](\sigma)
=
\frac{1}{2^{2k}(2k)!}
\int_{\sigma}^{\infty}
(E^{2}-\sigma^{2})^{2k}
f_{\text{therm}}(E)\, dE.
\end{equation}
Therefore,
\begin{equation}
\boxed{
I_{\text{therm}}^{(4k+3)}(\sigma)
=C_{4k+3}
\, 2^{2k} (2k)!
\, \mathcal{I}^{2k}[I_0](\sigma)
=\mathcal{A}\mathcal{I}^{2k}[I_0](\sigma)}
\end{equation}
and in the same way for $n=4k$ or since $\frac{d'-3}{2}=4k$ for $d'=8k+3$ we take
\begin{equation}
\boxed{
I_{\text{therm}}^{(8k+3)}(\sigma)
=C_{8k+3}
\, 2^{4k} (4k)!
\, \mathcal{I}^{4k}[I_0](\sigma)
=\mathcal{B}\mathcal{I}^{4k}[I_0](\sigma)}.
\end{equation}
where $\mathcal{A}$ and $\mathcal{B}$ are constants depending on $k$ index.

\section{Conclusion}

We have uncovered a deep mathematical structure in the gap equations of odd-dimensional Gross--Neveu models: the finite part follows higher-order discrete Laplacian patterns, while the cutoff-dependent part forms a truncated asymptotic expansion of a hypergeometric function. This structure appears only in $d = 4k+3$ dimensions and suggests these theories are secretly equivalent to $1d$ quantum mechanical problems with exact solvability properties. The discovery opens several directions including generalization to other models, connection to integrability, implications for the phase structure of thermal field theories in odd dimensions but also for thermal bootstrap methods and inversion formulas where transcendental mass-fixing conditions appear.

\subsection*{Acknowledgements} I would like to thank Anastasios Petkou for helpful discussion.

\subsection*{Data} All data that support the findings of this study are included within the article.

\subsection*{Conflict of Interest Statement}
The author declares no conflict of interest.

\subsection*{Funding Statement}
This research received no external funding.

\end{document}